\input{aipcheck}

\documentclass[
    ,final            
  ]
  {aipproc}

\layoutstyle{6x9}

\begin{document}

\title{A near field cosmology study of heavy elements in very metal-poor stars}


\author{C. J. Hansen}{
  address={Landessternwarte, ZAH, K\"onigstuhl 12, 69117 Heidelberg, Germany}
}



\begin{abstract}
Studying a range of old metal-poor stars provides information over cosmological timescales of our Galaxy. 
Such studies are indicative of the pristine gases and evolution of the Milky Way. 
Deriving stellar parameters and abundances from high-resolution observations of stars at various stellar evolution stages (including old dwarfs and RR Lyrae), allows us to use these abundances as tracers of an even earlier progenitor population.
Here, we carry out a detailed abundance study of mainly heavy elements (Z $>$ 38), i.e. neutron-capture elements, which we at low metallicities ([Fe/H] $< -2.5$) take as pure supernova type II products. A comparison of the derived abundances to type II supernova yields of heavy elements provides knowledge of the old stellar generations as well as properties of neutron-capture formation sites.
\end{abstract}

\maketitle


\section{Introduction}
Despite the efforts made to understand and constrain the first stars and the process of element formation connected to them, we are after several decades still left with many unanswered questions. How massive were the first stars, and how were the heavy elements we detect in old metal-poor stars created? One way of answering these questions is to study old metal-poor stars. From these stars we can derive accurate stellar abundances, allowing us to trace the progenitor stars that yielded the gas the later stellar generations were made from. It is therefore important to understand if this gas was pure, or mixed with other gases in the interstellar medium. 
Most old dwarf stars of spectral type F and G preserve the original gas composition in their outer atmospheres. This allows us to study pristine gases of elements between lithium and uranium. The lighter elements (Z $< 26$) provide information on the characteristics of the supernova that created them during the explosion. The heavy elements (Z $> 38$) are formed by neutron-capture (n-capture) processes. We have detailed information for some n-capture processes while others remain poorly constrained. By studying abundances of elements that we can tie to the different processes, we can learn about their similarities and differences, which will bring us closer to understanding the formation processes as well as the supernovae (SNe) that host them. 
 
\section{Data, Sample and Method}
The data cover 73 stars in total, two RR lyrae, 29 giants and 42 dwarfs. The high-resolution spectra have been obtained with UVES/VLT \cite{dekker} and Hires/Keck \cite{vogt94}. The data reduction and stellar parameter determination is described in \cite{CJHag}.
To derive the abundances we used the 1D local thermodynamic equilibrium (1D LTE) spectral synthesis code MOOG \cite{snedenphd} and MARCS model atmospheres \cite{Gus08}.

\section{Tracing the first supernovae with RR Lyrae stars}
\begin{figure}
  \includegraphics[height=.45\textheight]{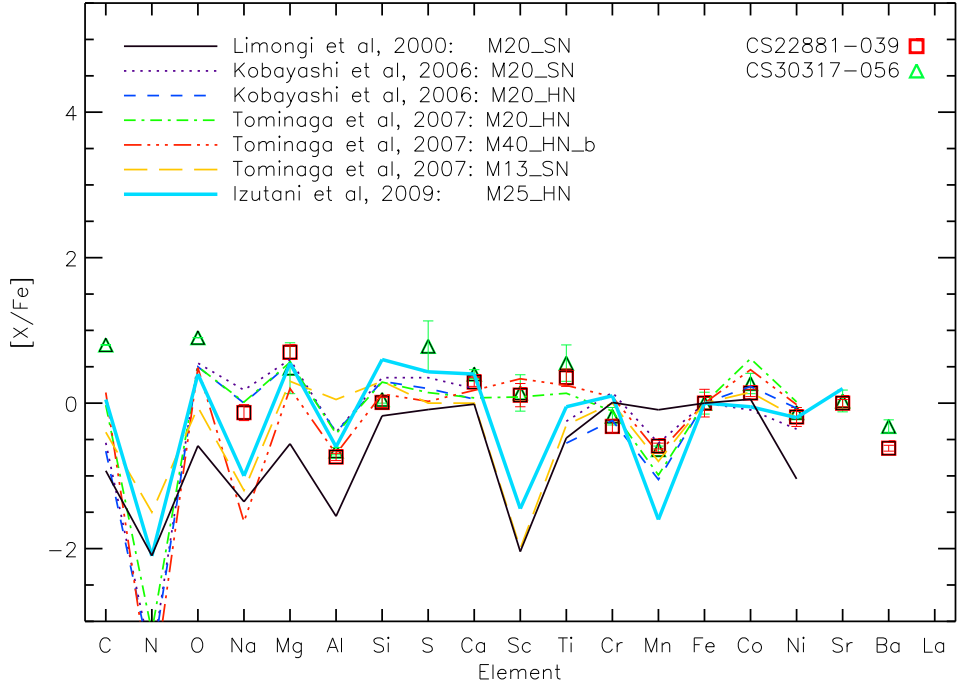}
  \caption{Stellar abundances derived for the two RR lyrae stars are compared to the SN yield predictions (references are listed in the legend).\label{SN}}
\end{figure}

Several studies have shown that very evolved stars, such as RR lyrae stars, can be used as chemical tracers \cite{taut,hansenRR,for} as they preserve the gases of elements heavier than lithium in their surfaces. 
By comparing the derived stellar abundances to a variety of SN yields we can extract information about the early (first?) SN such as mass, energy and electron fractions ($Y_e$). This may provide insight into the nature of the first stars. We compare to yield predictions from \cite{lim,koba,tom,Izu} covering a mass range of 13-40$M_{\odot}$, and an energy range of 1-10 foe, with or without jets/asymmetric explosions (see Fig. \ref{SN}).
The odd-even effect help to constrain the SN progenitor mass, the iron-peak elements are connected to the peak temperature of the explosion, while e.g. Sc provides information on the explosion energy and $Y_e$. The comparison in Fig. \ref{SN} points towards an early generation of SN, which existed well below [Fe/H]=$-3.3$ with a progenitor mass of around 40$M_{\odot}$ and high explosion energy. However, several quantities such as the estimated SN mass will be lowered if non-LTE effects are considered. These SN could be the objects that created the very heavy elements during their explosion. 

\begin{figure}
  \includegraphics[height=.28\textheight]{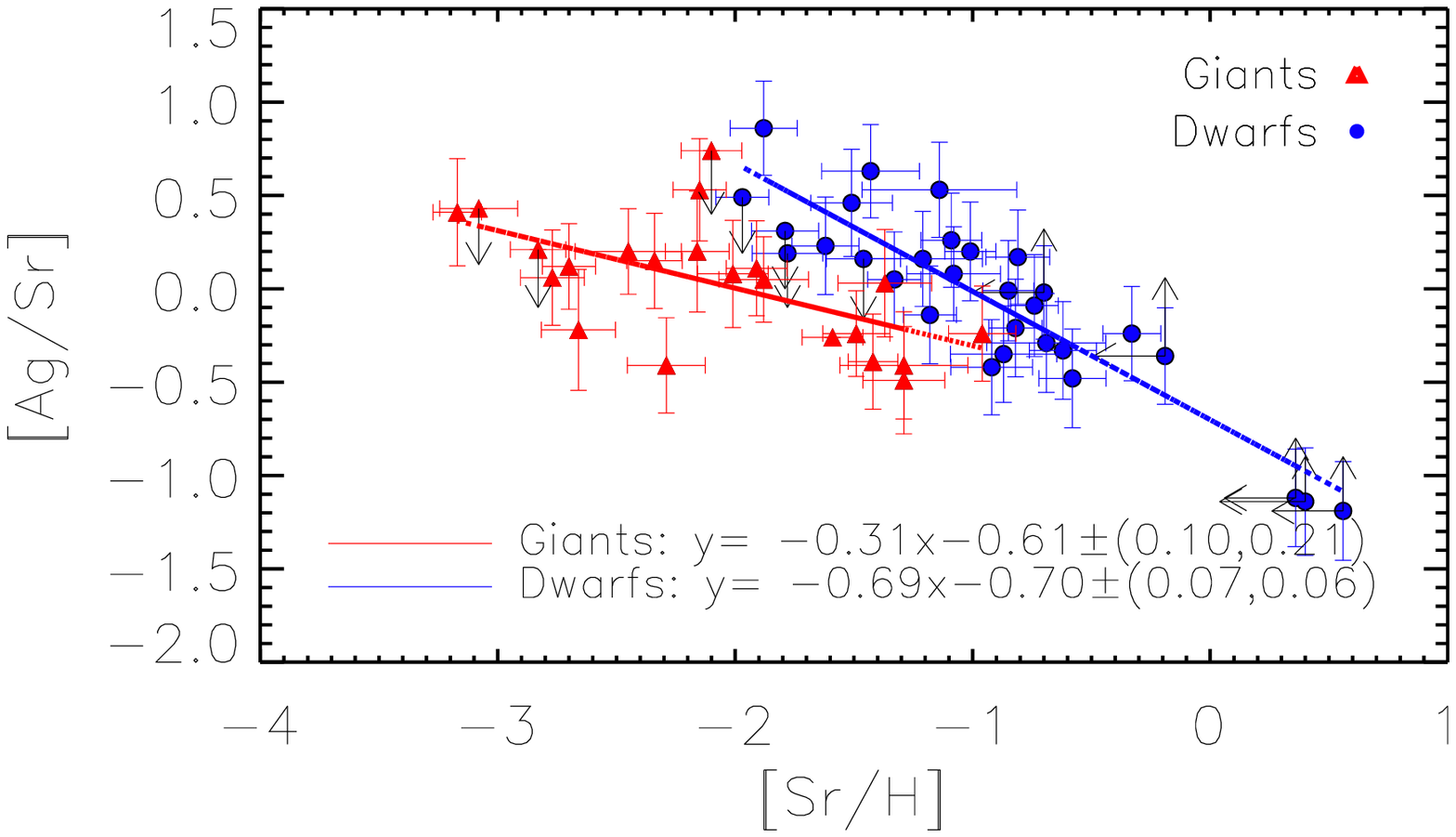}
\end{figure}
\vspace{-19mm}
\begin{figure}
  \includegraphics[height=.28\textheight]{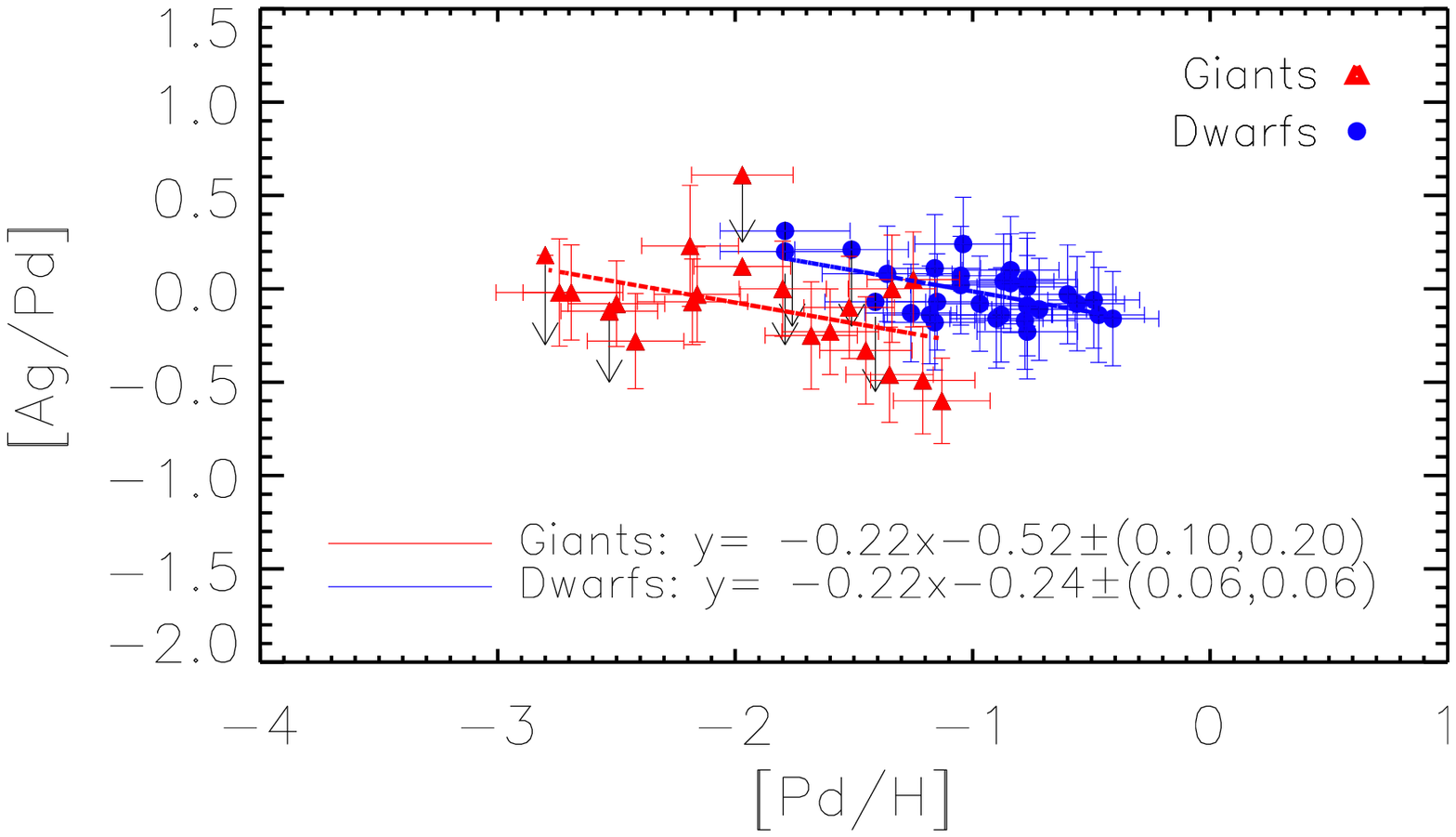}
\end{figure}
\vspace{-9mm}
\begin{figure}
  \includegraphics[height=.32\textheight]{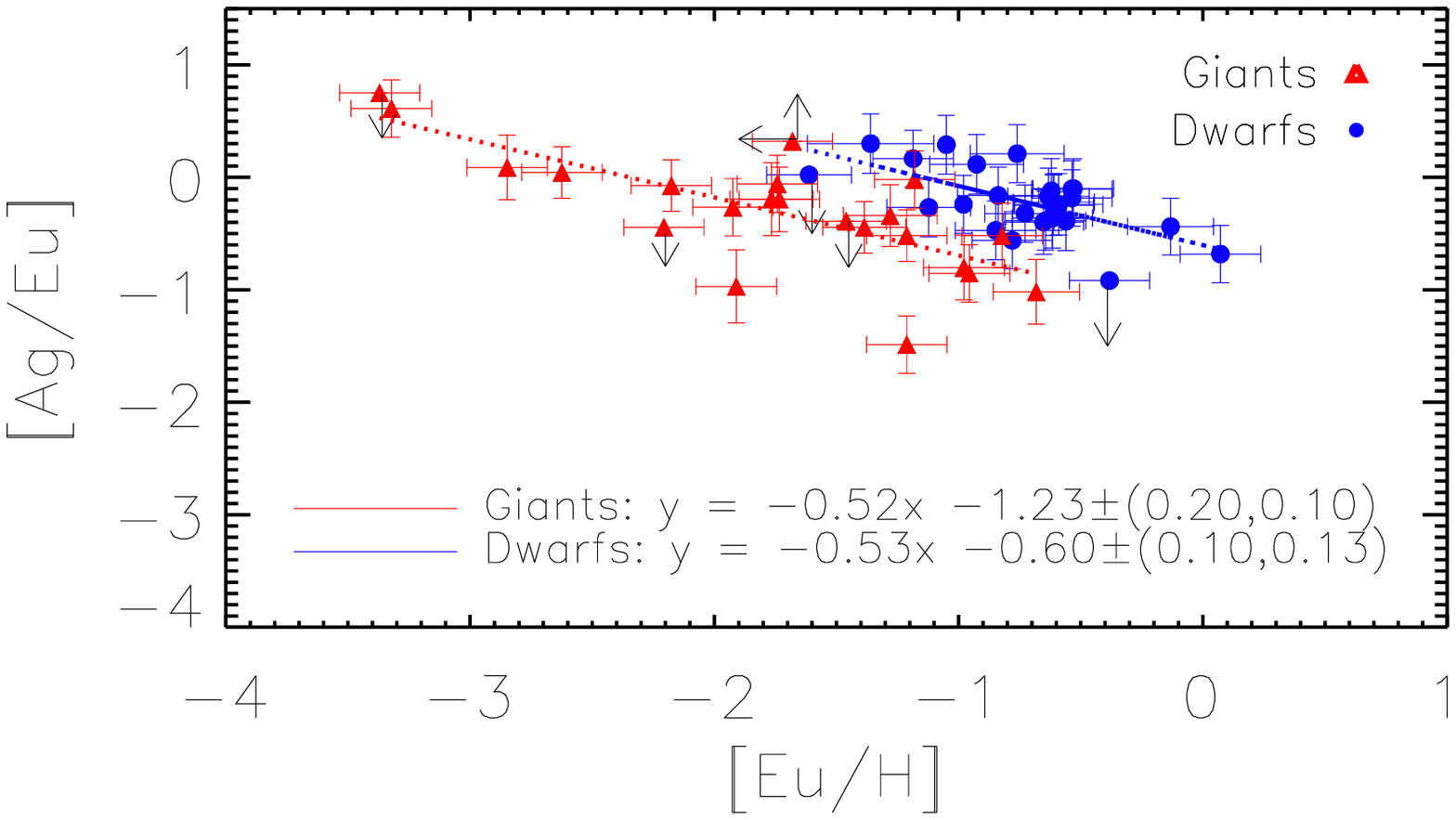}
  \caption{Silver compared to Sr, Pd, and Eu in the giant and dwarf stars from our sample.\label{SrAgEu}}
\end{figure}
\clearpage

\section{The evolution of heavy elements}
The majority of the heavy elements are produced in neutron-rich, very energetic environments. 
There are two main channels through which the heavy elements can be made: a secondary slow n-capture (s-)process and a primary rapid n-capture (r-)process. Each of these production channels seem to branch into two (weak and main) n-capture processes. Here an observational study is presented attempting to map the features of the weak r-process.

The elements that were targeted for this study are: Sr, Y, Zr, Pd, Ag, Ba, and Eu. Pd and Ag are assumed to be created by the weak r-process, whereas e.g. Sr is created by a weak s-process \cite{heil} (or charged particle process at low metallicity) and Eu is formed by a main r-process \cite{arland} at all metallicities. In order to understand the weak r-process better we compare Ag to Sr and Pd to see how their formation processes differ. If the two elements are formed by the same process this will be seen as a flat (horizontal) trend in the abundance plots, otherwise the trend will be a line with a negative slope.
Figure \ref{SrAgEu} shows that Pd and Ag share a formation process, which is very different from the weak s-process/charged particle process and the main r-process. This weak r-process is also very different from the main s-process responsible for creating Ba (cf. \cite{CJHag}).
To understand the characteristics and environment of the weak r-process, the observationally derived abundances are compared to self-consistent faint O-Ne-Mg electron-capture SN 2D models \cite{wan11} as well as the high-entropy wind (HEW) \cite{farou} parameter studies.    

\begin{figure}[!h]
  \includegraphics[height=.25\textheight]{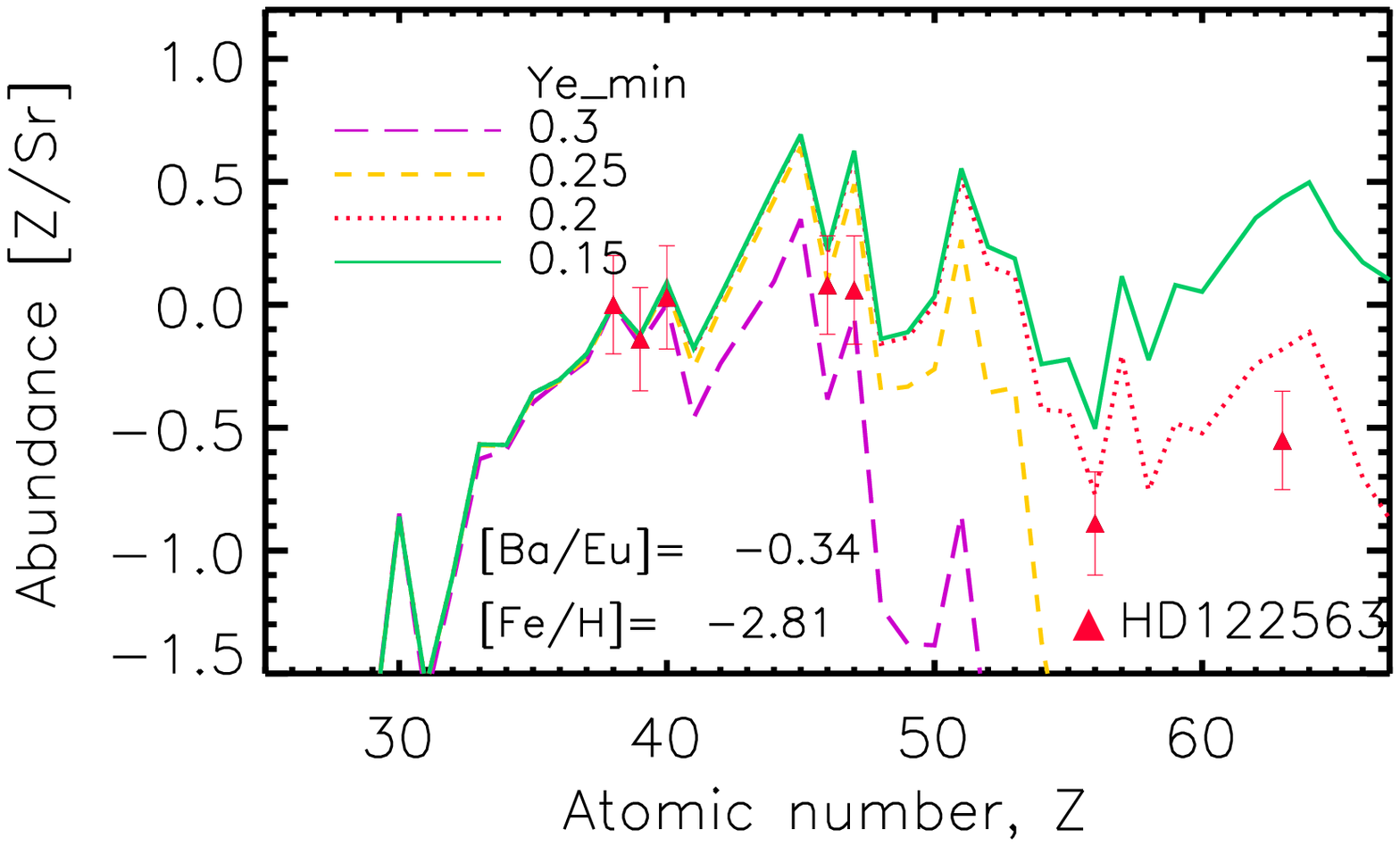}
  \includegraphics[height=.24\textheight]{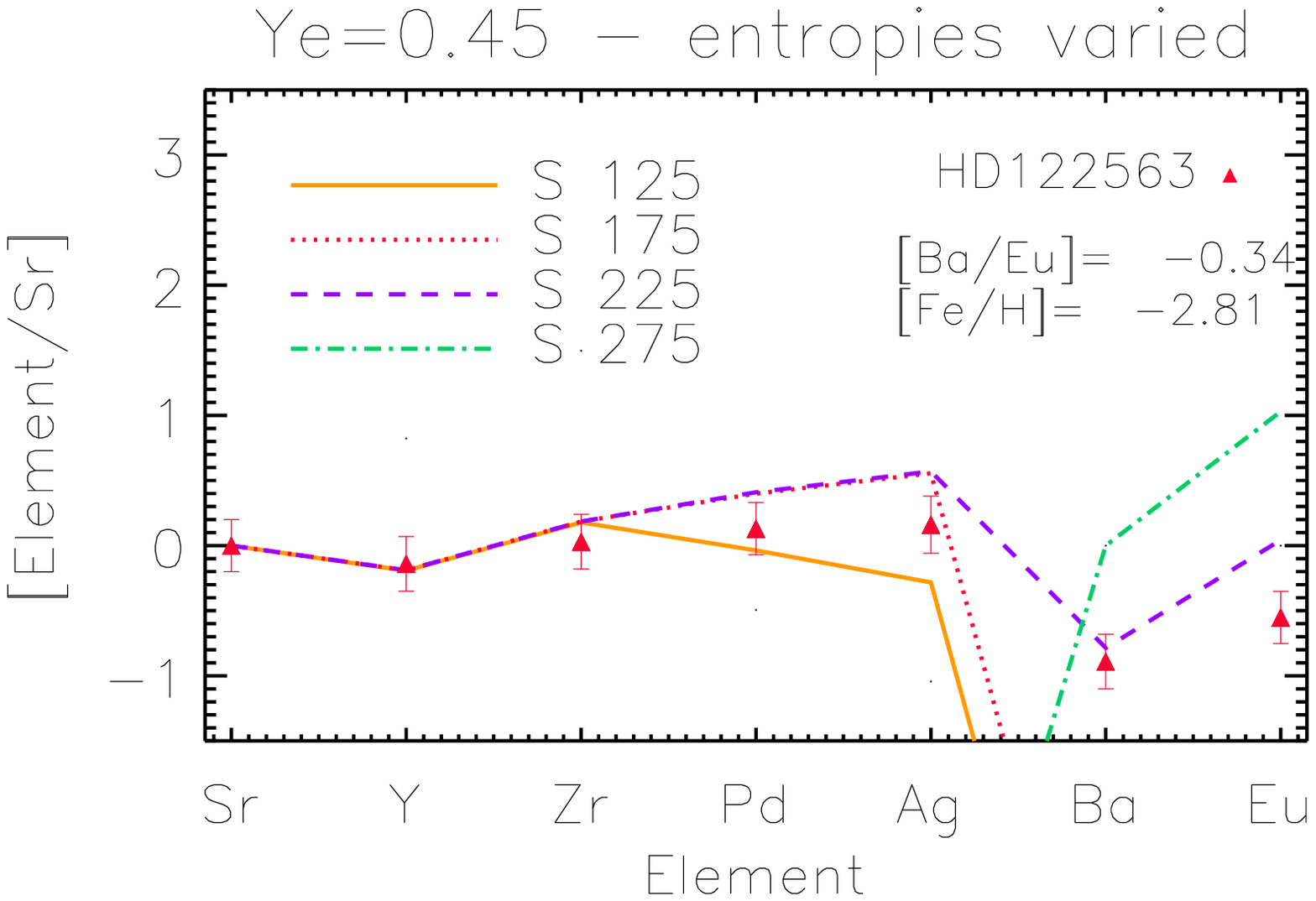}
  \caption{Abundances of all seven elements compared to model yield predictions. Three stars have been selected, one that shows a pure r-process pattern, one with an r-poor, and one with an r-rich abundance pattern. The stars are compared to O-Ne-Mg SN [16] on the left-hand side and to the HEW predictions on the right-hand side.\label{model}}
\end{figure}

The right-hand plots in Fig. \ref{model} shows the abundances compared to the HEW model predictions, and the left-hand plots illustrates the O-Ne-Mg SN yields. In general Sr-Ag can in all stars and both models be produced by one model/one set of parameters, while Ba and Eu require very different parameters/features in the formation processes. The elements lighter than Ag can be correctly predicted either by an entropy $125 \leq S \leq 175 k_B/baryon$ or an $Y_e \geq 0.25$, while Ba and Eu generally need $S > 225$ and $Y_e < 0.2$. This is also true for stars with pure r-process abundance patterns ([Ba/Eu]$<-0.74$), which confirms the r-process nature. These results apply to both r-poor and r-rich stars. It is evident from the comparison to the two models that r-poor and r-rich stars cannot be produced at the same site. The faint O-Ne-Mg SN seems to be a promising site for the weak r-process, and the HEW models yield a pattern that fits those of the r-rich stars well.

\section{Summary}
The abundances derived under non-LTE, compared to those derived under the LTE assumption, will provide e.g. a different mass and energy of the SN we are trying to trace. Hence it is central to understand the abundances and stellar parameters when these are derived including non-LTE and 3D effects.

\begin{figure}[!h]
  \includegraphics[height=.25\textheight]{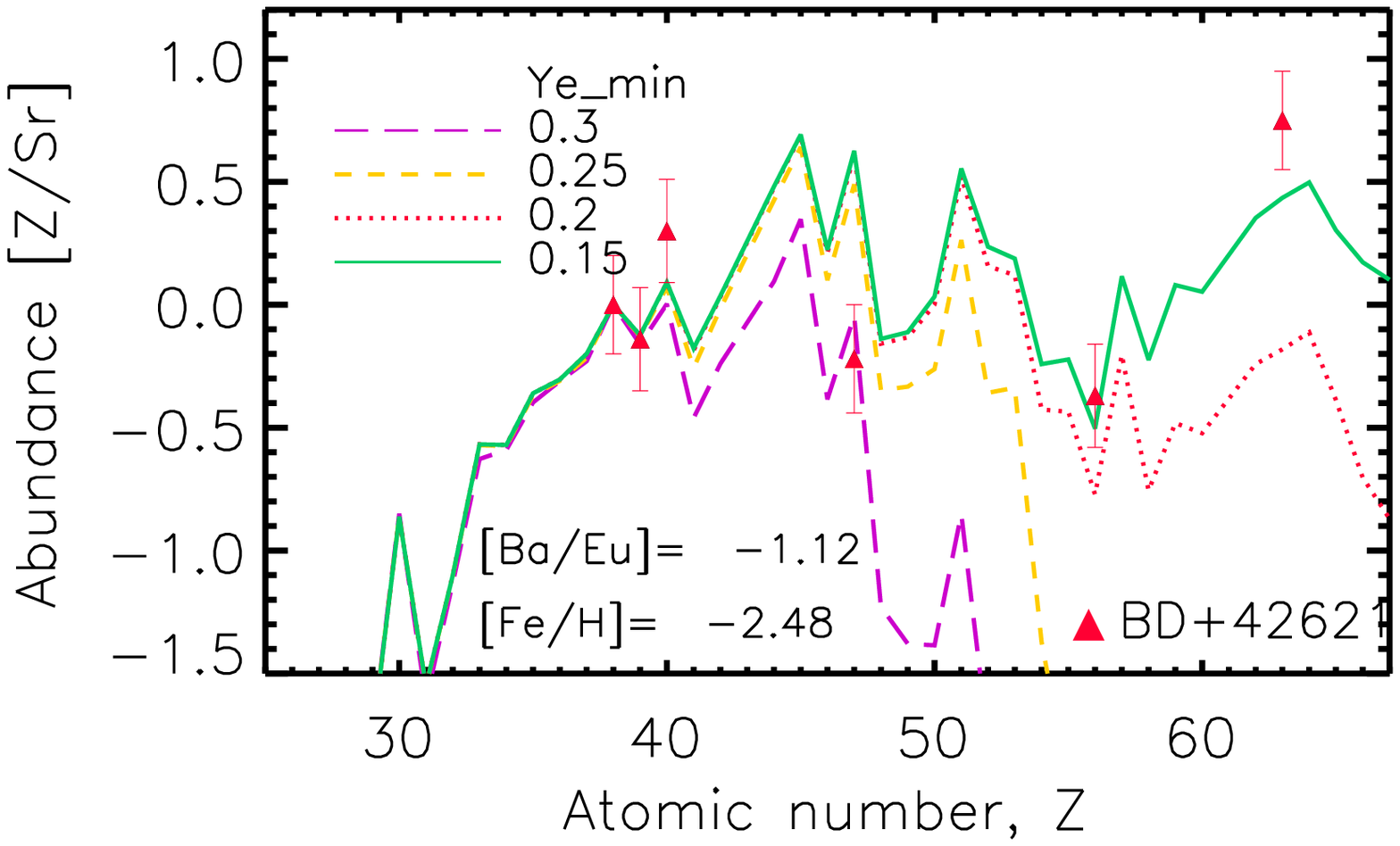}
  \includegraphics[height=.24\textheight]{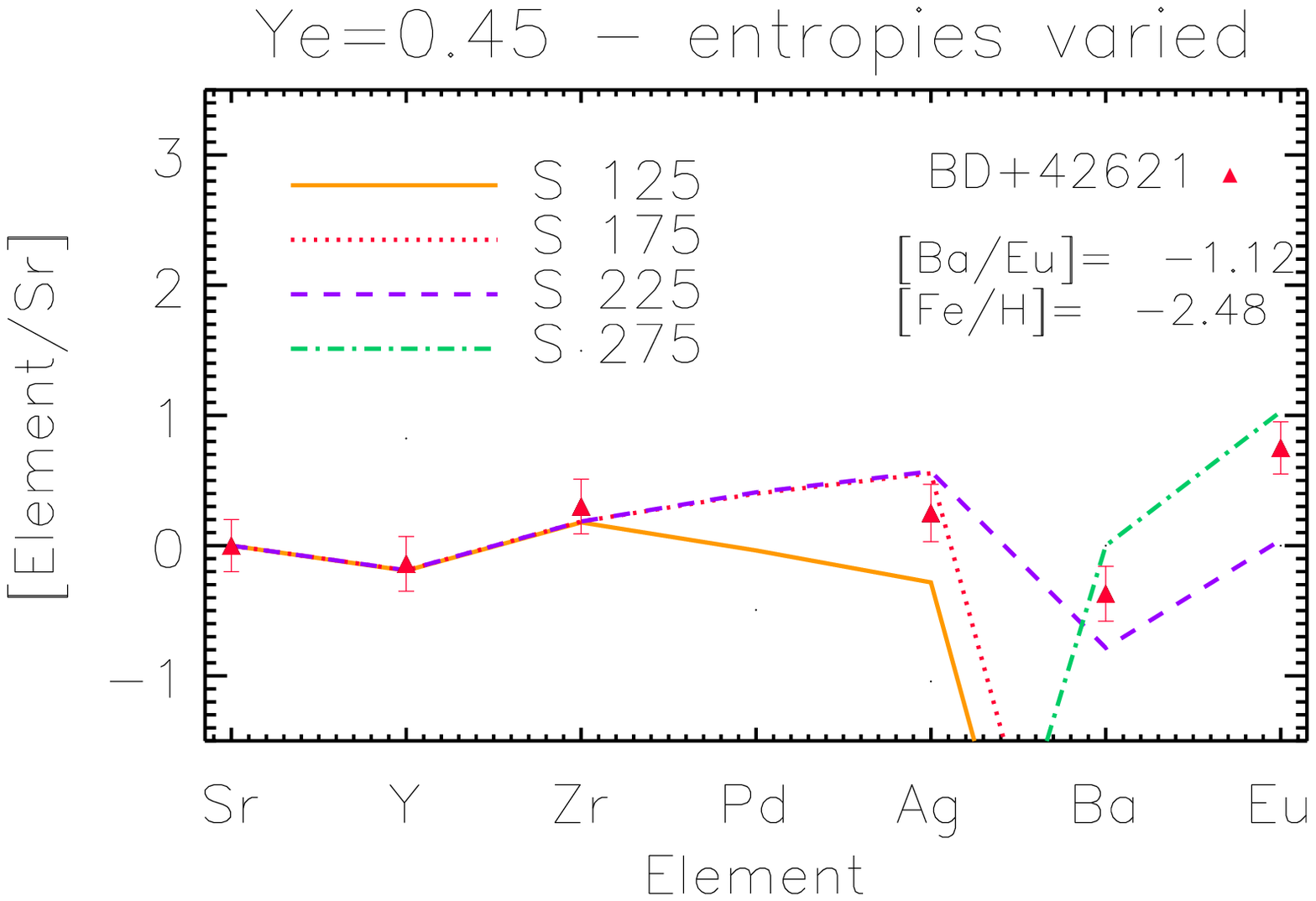}
\end{figure}
\begin{figure}[!h]
  \includegraphics[height=.25\textheight]{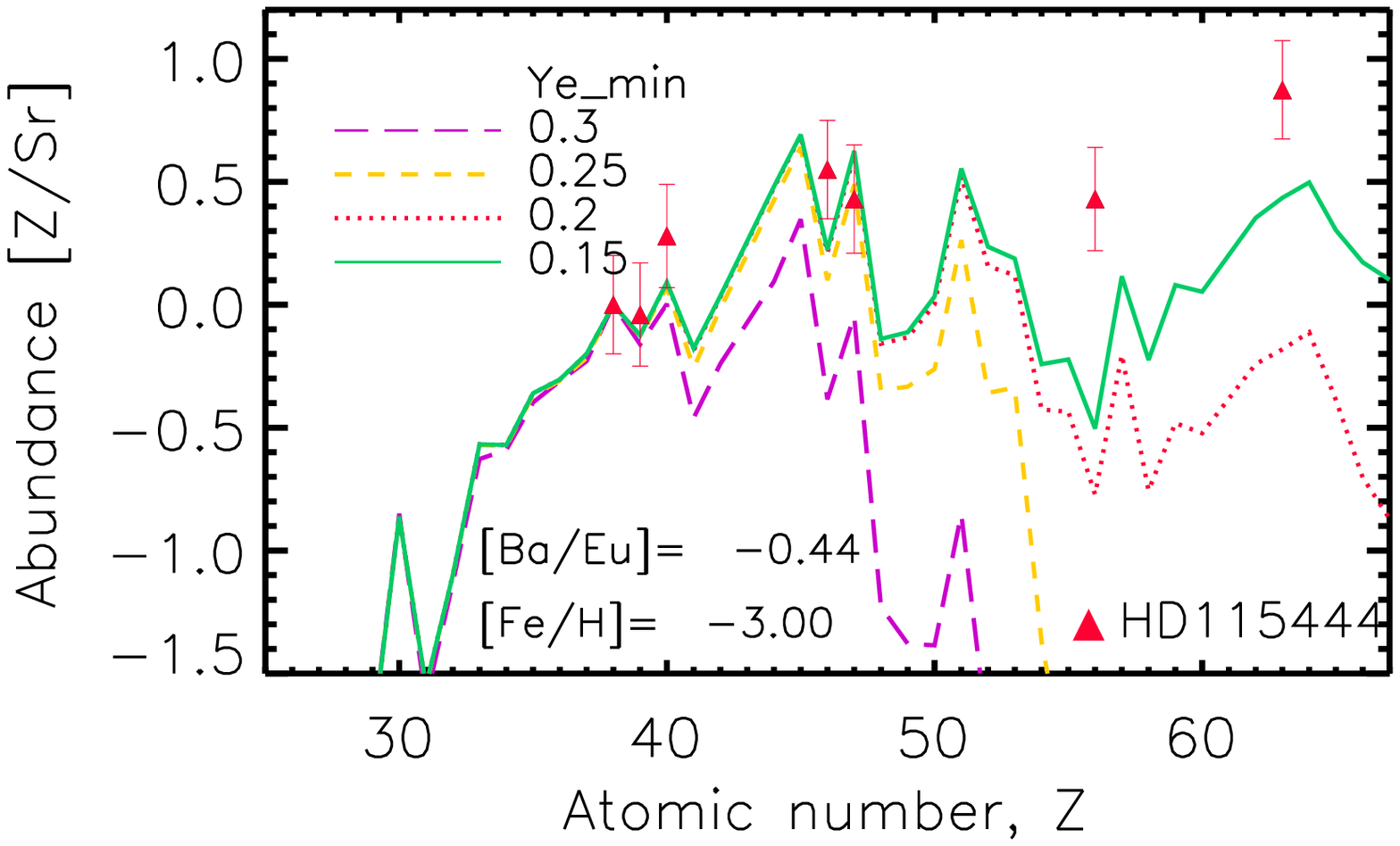}
  \includegraphics[height=.24\textheight]{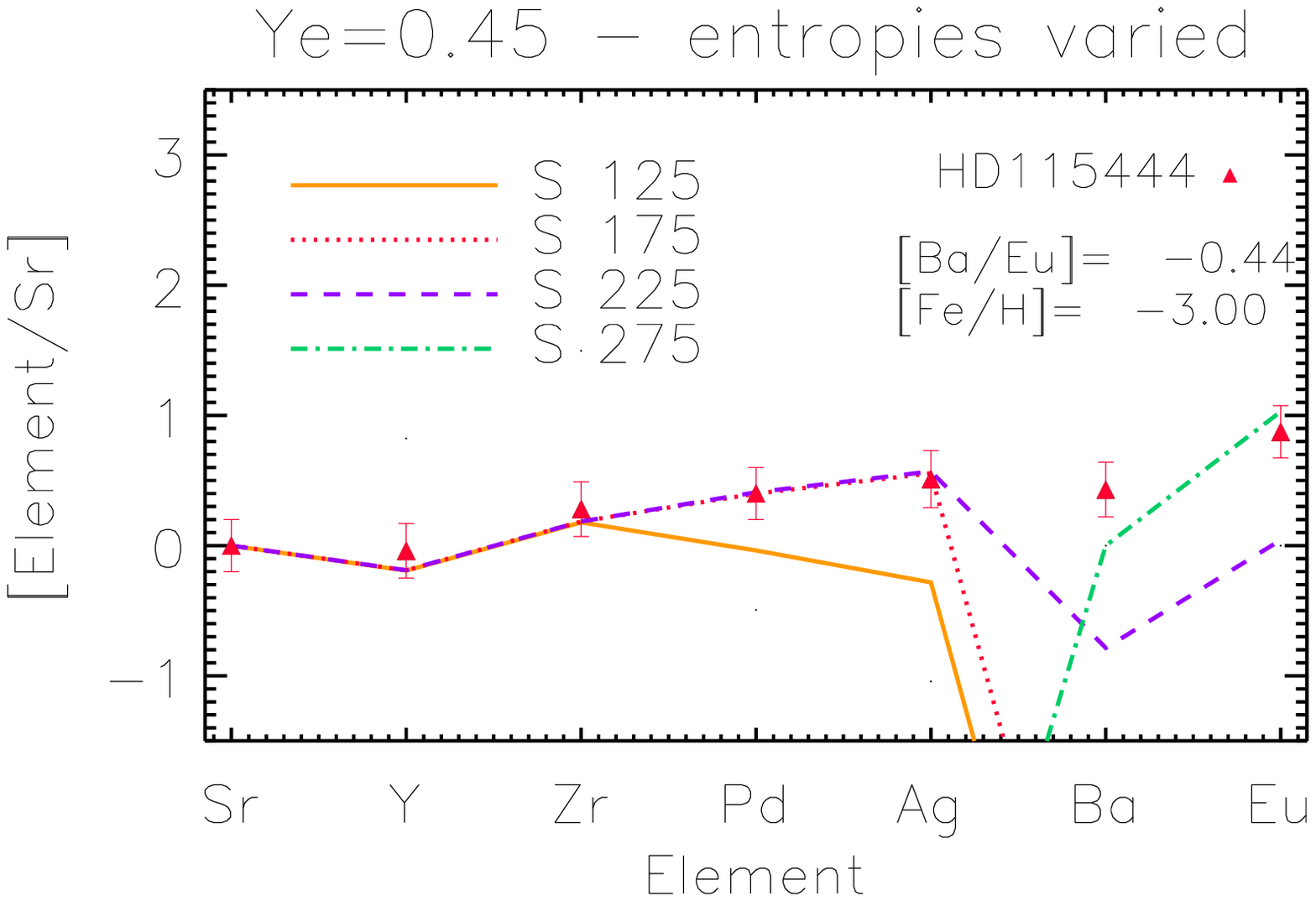}
\end{figure}

To explain Pd and Ag abundances we need a weak r-process, which so far seems to differ strongly from the weak/main s-process as well as the main r-process. The weak r-process is efficiently yielding Pd and Ag at metallicities below [Fe/H]$=-3.3$. This weak r-process needs lower neutron densities as well as entropies compared to what the main r-process needs, but higher such values than the s-process requires. We need 3D self-consistent SN models and optimised networks to understand quantities such as entropy and electron fractions before we can constrain this weak r-process.   


\begin{theacknowledgments}
  This work was supported by Sonderforschungsbereich SFB 881 'The Milky Way System' (subproject A5) of the German Research Foundation (DFG). CJH is grateful to F. Primas, H. Hartman, K.-L. Kratz, S. Wanajo, B. Leibundgut, K. Farouqi, N. Christlieb, B. Nordstr\"om, P. Bonifacio, M. Spite, J. Andersen, and the First Stars IV organizers.
\end{theacknowledgments}



\bibliographystyle{aipproc}   


\begin{thebibliography}{0}
\expandafter\ifx\csname natexlab\endcsname\relax\def\natexlab#1{#1}\fi
\providecommand{\enquote}[1]{``#1''}
\expandafter\ifx\csname url\endcsname\relax
  \def\url#1{\texttt{#1}}\fi
\expandafter\ifx\csname urlprefix\endcsname\relax\def\urlprefix{URL }\fi
\providecommand{\eprint}[2][]{\url{#2}}

\end{thebibliography}


\begin{thebibliography}{16}

\bibitem[{Dekker} et~al.(2000)]{dekker}
H.~{Dekker}, S.~{D'Odorico}, A.~{Kaufer}, et al. 2000, Proc. SPIE, 4008, 534.

\bibitem[{Vogt} et~al.(1994)]{vogt94}
S.~S. {Vogt}, S.~L. {Allen}, B.~C. {Bigelow}, et al. 1994, Proc. SPIE, vol.
  2198, 362.

\bibitem[{Hansen} et~al.(2012)]{CJHag}
C.~J. {Hansen}, F.~{Primas}, and H. {Hartman}, et al. 2012, A\&A, accepted.

\bibitem[{Sneden}(1973)]{snedenphd}
C.~A. {Sneden}, Ph.D. thesis, The University of Texas at Austin (1973).

\bibitem[{Gustafsson} et~al.(2008)]{Gus08}
B.~{Gustafsson}, B.~{Edvardsson}, K.~{Eriksson},et al. 2008, A\&A, 486, 951.

\bibitem[{Tautvaisiene}(1997)]{taut}
G.~{Tautvaisiene}, 1997, MNRAS, 286, 948.

\bibitem[{Hansen} et~al.(2011)]{hansenRR}
C.~J. {Hansen}, B.~{Nordstr{\"o}m}, P.~{Bonifacio}, et al. 2011, A\&A, 527, A65.

\bibitem[{For} et~al.(2011)]{for}
B.-Q. {For}, C.~{Sneden}, and G.~W. {Preston}, 2011, APJs, 197, 29.

\bibitem[{Limongi} et~al.(2000)]{lim}
M.~{Limongi}, O.~{Straniero}, and A.~{Chieffi}, 2000, APJs, 129,
  625.

\bibitem[{Kobayashi} et~al.(2006)]{koba}
C.~{Kobayashi}, H.~{Umeda}, K.~{Nomoto}, et al. 2006, APJ, 653, 1145.

\bibitem[{Tominaga} et~al.(2007)]{tom}
N.~{Tominaga}, H.~{Umeda}, and K.~{Nomoto}, et al. 2007, APJ, 660, 516.

\bibitem[{Izutani} et~al.(2009)]{Izu}
N.~{Izutani}, H.~{Umeda}, and N.~{Tominaga}, 2009, APJ , 692, 1517.

\bibitem[{Heil} et~al.(2009)]{heil}
M.~{Heil}, A.~{Juseviciute}, F.~{K{\"a}ppeler}, et al. 2009, PASA, 26, 243.

\bibitem[{Arlandini} et~al.(1999)]{arland}
C.~{Arlandini}, F.~{K{\"a}ppeler}, K.~{Wisshak}, et al. 1999, APJ , 525, 886.

\bibitem[{Wanajo} et~al.(2011)]{wan11}
S.~{Wanajo}, H.-T. {Janka}, and S.~{Kubono},  2011, APJ, 729, 46.


\bibitem[{Farouqi} et~al.(2010)]{farou}
K.~{Farouqi}, K.~{Kratz}, B.~{Pfeiffer}, et al. 2010, APJ, 712, 1359.

\end{thebibliography}


\end{document}